\documentclass[amsmath,amssymb,amsbsy,prl,twocolumn,tightenlines,superscriptaddress,floatfix]{revtex4-1}
\usepackage{amsfonts} 
\usepackage{amsmath} 
\usepackage{amssymb}
\usepackage{physics}
\usepackage{graphicx, nicefrac}
\usepackage{blindtext}
\usepackage{lineno}
\usepackage{cancel}
\usepackage{scalerel}
\usepackage[normalem]{ulem}
\usepackage{xfrac}
\usepackage{bm}
\usepackage[usenames, dvipsnames]{color} 
\usepackage{dcolumn}
\usepackage{epic} 
\usepackage{epsfig}
\usepackage{eucal} 
\usepackage{esdiff}
\usepackage{graphicx}
\usepackage[breaklinks,colorlinks = true,linkcolor = magenta,urlcolor=magenta,citecolor=red]{hyperref}
\usepackage{mathrsfs}
\usepackage{soul}
\usepackage{subfigure}
\usepackage{textcomp}
\usepackage{times}
\usepackage{verbatim}
\usepackage{wrapfig} 
\usepackage{xy} 
\usepackage{lineno}
\usepackage{float}
\usepackage{import}

\begin{document}
\title{Onset of Intermittency \textcolor{black}{and Multiscaling} in Active Turbulence}
\author{Kolluru Venkata Kiran} 
\email{kiran8147@gmail.com}
\affiliation{Centre for Condensed Matter Theory, Department of Physics, Indian Institute of Science, Bengaluru, 560012, India}
\author{Kunal Kumar} 
\email{kunal.kumar@icts.res.in}
\affiliation{International Centre for Theoretical Sciences, Tata Institute of Fundamental Research, Bengaluru 560089, India}
\author{Anupam Gupta} 
\email{agupta@phy.iith.ac.in}
\affiliation{Department of Physics, Indian Institute of Technology (IIT), Hyderabad, Kandi Sangareddy Telangana, 502285, India}
\author{Rahul Pandit} 
\email{rahul@iisc.ac.in}
\affiliation{Centre for Condensed Matter Theory, Department of Physics, Indian Institute of Science, Bengaluru, 560012, India}
\author{Samriddhi Sankar Ray} 
\email{samriddhisankarray@gmail.com}
\affiliation{International Centre for Theoretical Sciences, Tata Institute of Fundamental Research, Bengaluru 560089, India}
\begin{abstract} 
	
	Recent results suggest that  highly active, chaotic, non-equilibrium
	states of \textit{living} fluids might share much in common with high
	Reynolds number, inertial turbulence. We now show, by using a
	hydrodynamical model, the onset of intermittency and the consequent
	multiscaling of Eulerian and Lagrangian structure functions as a
	function of the bacterial activity.  Our results bridge the worlds of
	low and high Reynolds number flows as well as open up intriguing
	possibilities of what makes flows intermittent.
	
\end{abstract}
\date{}
\maketitle

The phenomenon of intermittency is most commonly associated with
high-Reynolds-number fluid flows that are turbulent~\cite{Frisch-CUP}. An
analytical theory of such intermittency has remained elusive because of the
formidable challenges posed by the nonlinearity of the Navier-Stokes equations
of hydrodynamics
\footnote{An analytical theory has been developed for Kraichnan's model of
passive-scalar turbulence~\cite{RMP-Grisha}, in which the passive-scalar
equation is \textit{linear}.}.  Nevertheless, experiments, observations and
numerical simulations of three-dimensional (3D) fluid turbulence (henceforth
inertial turbulence) clearly show strong evidence of
intermittency~\cite{Frisch-CUP}, which manifests itself, \textit{inter alia},
in the measured deviations of the exponent $\zeta_p$ of the order-$p$,
equal-time, inertial-range velocity structure functions from the mean-field,
dimensional linear $p/3$ result of Kolmogorov~\cite{kolmogorov1991local} from
1941 (K41). Specifically, $\zeta_p \ge p/3$ for $p < 3$ and $\zeta_p \le p/3$
for $p \ge 3$ with $\zeta_3 = 1$~\cite{Frisch-CUP}. Such intermittency or
anomalous scaling has also been seen in magnetohydrodynamic~\cite{Basu-MHD},
passive-scalar (advected by the Navier-Stokes velocity)~\cite{Mitra-Pandit2005}
and quantum~\cite{Pomyalov-superfluid,rusaouen2017pof,verma2023statistical} turbulence.
\textcolor{black}{Consequently, in the fluid-turbulence community, a perception has emerged that intermittency and 
anomalous scaling are a consequence of the high-Reynolds-number fluid turbulence.
However, recent results in low-Reynolds systems~\cite{liu2012multifractal,xang2016intermittency,nadia2023prr,Rahul1,Rahul2,romaguera2024multifractal,huang2024logbact}, suggest that this is not entirely correct.}

\textcolor{black}{A starting point is a reexamination of this perception} in
dense bacterial suspensions which behave like (living) fluids at approximately
zero Reynolds numbers\textcolor{black}{
~\cite{dombrowski2004self,wensink2012meso,dunkel2013fluid,marchetti2013hydrodynamics,ramaswamy2017active,
alert2021}.} 
The emergent fluidised state of such dense suspensions is known to
display a variety of dynamical
phases~\cite{james2021emergence,Sid21,puggioni2022giant}. Most prominently, for
a range of activity $\alpha$, these two-dimensional, low-Reynolds number fluids
are marked by chaotic, vortical motion, which is remarkably similar to inertial
high-Re turbulence~\cite{BoffettaEcke,Pandit-2DReview}. Hence, the nomenclature
\textit{active turbulence}~\cite{wensink2012meso} to describe such
non-equilibrium states~\cite{alert2021,bhattacharjee2022activity}.  Not
surprisingly, active turbulence has been the subject of several recent studies,
some with biological motivations, such as evasion and
foraging~\citep{humphries2010environmental,volpe2017topography}, and other
investigations of the nature of active turbulence and its relation with fluid
turbulence~\cite{bratanov2015new,zanger2015analysis,james2018vortex,james2021emergence,joy2020friction,Sid21,kiran2022irreversiblity,Sid22,Gibbon2023}.

\textcolor{black}{There is an additional important reason for us to study this system. The question of multiscaling 
in active turbulence was examined experimentally a decade ago with contradictory conclusions: The experiments of 
Liu and I~\cite{liu2012multifractal} found evidence for the multiscaling of $\zeta_p$; by contrast, Wensink \textit{et al.}~\cite{wensink2012meso} 
reported a complete absence of such signatures of intermittency. We resolve this contradiction by showing how active turbulence
can display \textit{a transition from simple scaling to multiscaling as a function of the activity parameter}. This 
offers a natural explanation for why experiments can lead to varying conclusions
regarding intermittency.}

Dense, bacterial suspensions follow a hydrodynamic description~\cite{wensink2012meso,alert2020universal} 
\begin{eqnarray}
\partial_t {\bf u} + \lambda {\bf u}\cdot \nabla {\bf u}&=& -\nabla p- \Gamma_0\nabla^2 {\bf u} - \Gamma_2 \nabla^4 {\bf u}
-\nonumber \\ &-& (\alpha + \beta \vert {\bf u} \vert^2) {\bf u}\,,
\label{eq:genHyd}\\
\nabla \cdot {\bf u}&=& 0\,,
\label{eq:incomp}
\end{eqnarray}
where ${\bf u}({\bf x},t)$ is the coarse-grained, incompressible,  velocity
field of the motile bacterial (active) suspension, with the positive parameter
$\lambda$ defining pusher-type bacteria. The chaotic patterns arise from
instabilities that arise because
$\Gamma_0,\Gamma_2>0$~\cite{swift1977hydrodynamic,wensink2012meso,simha2002hydrodynamic,linkmann2020condensate,bratanov2015new}. Finally, the Toner-Tu drive~\cite{TT95,TT98} is accounted for by the terms with coefficients $\alpha$
and $\beta$. Stability demands $\beta > 0$; and the negative activity parameter
$\alpha$ injects energy; the more active
the suspension the more negative is $\alpha$. The TTSH equation~\eqref{eq:genHyd} has the same advective nonlinearity as the 
Navier-Stokes (NS) equation. Hence, 
it is tempting to ask if there is a range of activity $\alpha$, in which TTSH turbulence has statistical properties that are akin to those of inertial turbulence. In particular, does this activity-driven turbulence possess
(approximate) scale-invariance, in a power-law range with a universal spectral scaling
exponent, fluctuations, and intermittency that leads to multiscaling?

\begin{figure*}[!]
	\includegraphics[width=1.0\linewidth]{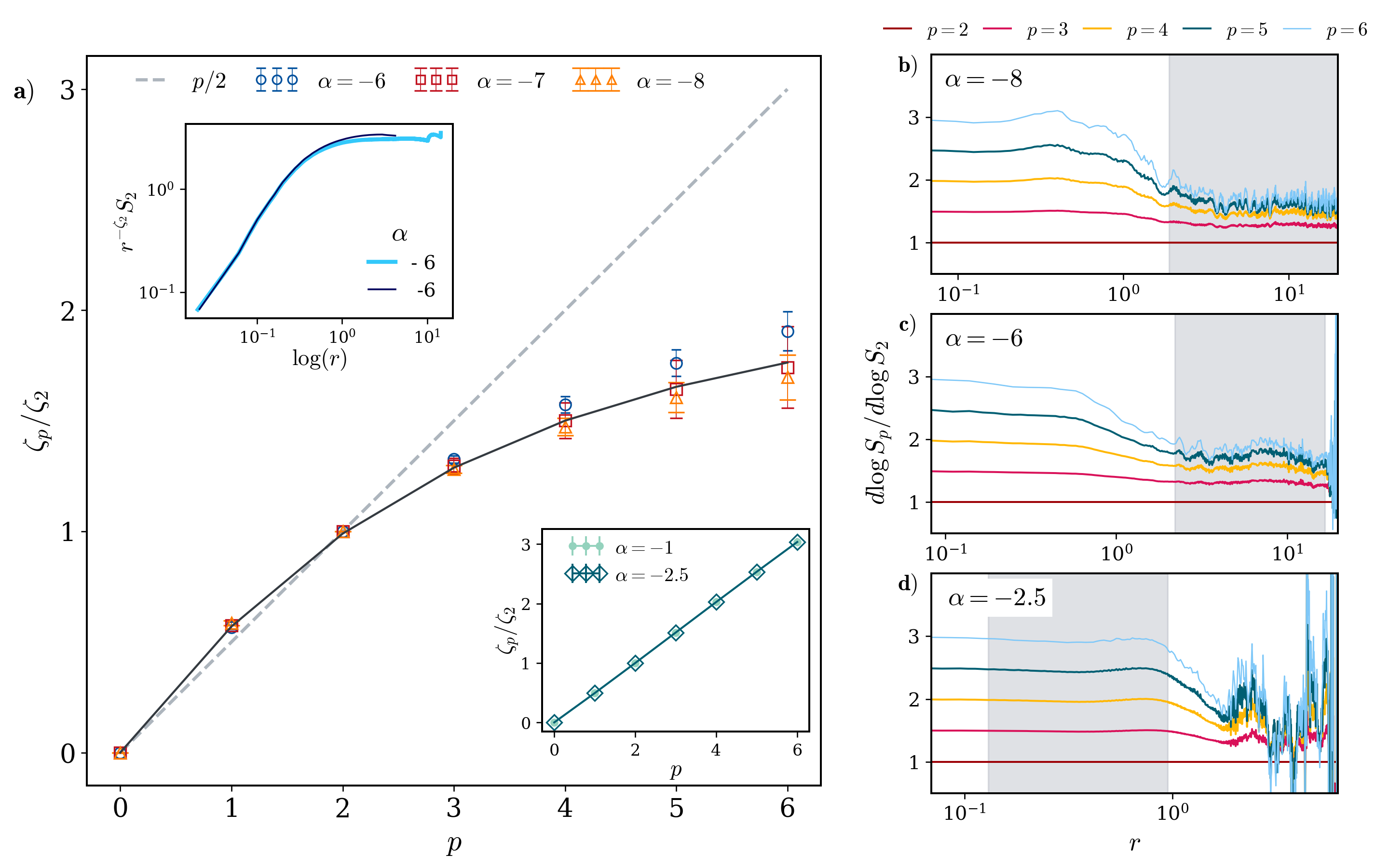}

	\caption{ (a) A plot of the exponent ratios $\zeta_p/\zeta_2$ (with error bars) obtained from an extended self-similarity (ESS) analysis of the 
	longitudinal structure function versus $p$ for different values of $\alpha \lesssim \alpha_c$ (see legend; for $\alpha = -6$ we show exponents from two different simulations with different domain sizes);
	the dashed black denotes the (linear) scaling $p/2$. In the upper 
	inset we show plots of the second-order (longitudinal) structure functions compensated with the scaling exponent $\zeta_2 = 0.5$ to show 
	our results are consistent with what has been reported earlier for the spectral scaling and the extent our scaling range. 
	In the lower inset, we show plots of $\zeta_p/\zeta_2$ versus $p$ for $\alpha \gtrsim \alpha_c$ and from both sets of simulations (see legend).
	Clearly, for such mildly active suspensions the scaling exponents are linear with the order $p$. (The error bars are consistent with symbol sizes and not shown explicitly for clarity.). \textcolor{black}{The local slopes $\tfrac{d \log S_p}{d \log r}$ vs $r$ for (b) $\alpha=-8$; (c) $\alpha=-6$ and (d) $\alpha=-2.5$. The shaded gray region shows the plateau 
	whose mean value yields the scaling exponents; error-bars follow from the standard deviations.}
	}
	\label{fig:zetap}
    \end{figure*}

These questions were partially answered in a recent paper by Mukherjee,
\textit{et al.}~\cite{SiddharthaNatPhys} who showed, through a closure analysis
complemented by direct numerical simulations, the existence of a critical
activity $\alpha_c$: For highly active suspensions $\alpha \lesssim \alpha_c$,
the energy spectrum has a universal activity-independent scaling exponent and
the distribution of velocity gradients and increments show distinct
non-Gaussian tails. For mildly active suspensions $\alpha \gtrsim \alpha_c$, in
contrast, the energy spectrum has a non-universal, activity-dependent
scaling~\cite{bratanov2015new} with Gaussian distributions of velocity
gradients and increments~\cite{wensink2012meso}. Furthermore, Kiran,
\textit{et al.}~\cite{kiran2022irreversiblity} have shown that probability distribution functions 
(PDFs) of temporal increments of the energy, along
trajectories of Lagrangian tracers, display wide, non-Gaussian tails.
However, the question of the possible multiscaling of
both Eulerian and Lagrangian velocity increments, the cornerstone of multiscaling in classical
turbulence theory [see, e.g., ~\cite{Frisch-CUP,Sreeni-saturation}], has remained unaddressed in 
active turbulence. We demonstrate that Eulerian and Lagrangian intermittency, 
in highly active TTSH turbulence, is similar to their inertial-turbulence counterparts. 

Our study uses pseudospectral direct numerical simulations of Eqs.~\eqref{eq:genHyd} on a doubly-periodic 
domain~\citep{james2018vortex}. We perform different sets of simulations, with domain sizes ranging from $L = 2\pi$ to 80,
and $1024^2$ to $8192^2$ collocation points.
We use $\Gamma_0 = 0.045$, $\Gamma_2 = \Gamma_0^3$, $\beta = 0.5$, 
$\lambda = 3.5$~\citep{wensink2012meso,james2018vortex,joy2020friction}, and the activity parameter $\alpha$ is varied 
across the transition $\alpha_c$. 




In analogy with inertial turbulence, we turn our attention to the Eulerian
scaling exponents $\zeta_p$ of the equal-time, longitudinal, order $p$
structure function $S_p (r) \equiv \langle |\Delta {\bf u}\cdot \hat {\bf
r}|^p\rangle \sim r^{\zeta_p}$, where $\Delta {\bf u} = {\bf u}({\bf x}+{\bf
r}) - {\bf u}({\bf x})$. In particular, the second-order exponent $\zeta_2$ is
trivially associated with the scaling exponent of the energy spectrum $E(k)
\sim k^{-\delta}$, for $1 \le \delta \le 3$, via the Wiener–Khinchin theorem,
as $\zeta_2 = \delta - 1$. For $\alpha \lesssim \alpha_c$,  we know $\delta = 3/2$~\cite{SiddharthaNatPhys} and thence $\zeta_2 = 1/2$.
We confirm this by showing, in the upper inset Fig.~\ref{fig:zetap} (a), the
compensated second order structure function $r^{-\zeta_2}S_2$ versus $r$ with
$\zeta_2$ set to 1/2 for $\alpha = -6 < \alpha_c$.  The flat plateau, at values
of $r$ consistent with the spectral scaling range~\cite{SiddharthaNatPhys}, is
a confirmation of the self-consistency in our measurements of the structure
functions. Furthermore, dimensional analysis suggests that, in the
absence of intermittency $\zeta_p = p/4$ (the analog of the K41
scaling $p/3$ in inertial turbulence). We now evaluate $S_p (r)$ for 
$p > 2$ to uncover the deviations, because of intermittency, from the
linear-scaling result $p/4$.


We measure high-order exponents in two ways: First, we use log-log plot of
$S_p$ vs. $r$ to obtain the exponent from their local slopes, i.e., $\zeta_p =
\langle \tfrac{d \log S_p}{d \log r} \rangle_r$. Here, $\langle \cdot
\rangle_r$ denotes the average over $r$, in the plateau in the upper inset of
Fig.~\ref{fig:zetap}(a), where the local slope is nearly flat. Second, we also
calculate the exponent ratio $\tfrac{\zeta_p}{\zeta_2}$ by using a local slope
analysis in conjunction with the extended-self-similarity (ESS)
procedure~\cite{BenziESS,RayESS}: $\tfrac{\zeta_p}{\zeta_2} = \langle \tfrac{d
\log S_p}{d \log S_2} \rangle_r$.  Both these methods yield consistent values
of $\zeta_p$; understandably, the error bars improve when we employ ESS, which
allows us to obtain a scaling range \textcolor{black}{of about a decade as
indicated in Figs.~\ref{fig:zetap}(b)-(d). The mean exponents are the average
values of the plateaus; the error-bars follow from the standard deviation.}

In Fig.~\ref{fig:zetap}(a) we show plots of $\tfrac{\zeta_p}{\zeta_2}$ versus
$p$, for different values of $\alpha \lesssim \alpha_c$, with a thick black
line passing through the mean values of these exponents, as a guide to the eye.
Within error bars, the exponents for the different values of $\alpha$ overlap;
we have checked that this is also true for different resolutions and system
sizes.  \textcolor{black}{Our measured exponent ratios show significant deviation from the simple
linear scaling. This deviation is, reminiscent of the convex, monotonically
increasing plots of equal-time exponents in inertial
turbulence~\cite{Frisch-CUP,BuariaPRL2023}, is the
first, theoretical evidence of multiscaling in active turbulence for values of
$\alpha$ beyond $\alpha_c$. It substantiates the experimental measurements made
by Liu and I~\cite{liu2012multifractal} as well as the suggestion of an
emergent intermittency in measurements of the PDFs of velocity increments and
the associated kurtosis~\cite{SiddharthaNatPhys}.} 

To study milder activities, we carry out DNSs with  $-1 \gtrsim \alpha  \gtrsim \alpha_c$. In this range of $\alpha$, the spectral scaling exponent is given by $\delta = \frac {\tau_{\rm eff}(2\alpha + 8\beta E_{\rm tot})}{\lambda} -
1\gtrsim 0$, with an effective time-scale $\tau_{\rm eff}$ and total energy
$E_{\rm tot}$ of the suspension~\cite{rose1978fully,BoffettaEcke,bratanov2015new,SiddharthaNatPhys}. Hence,
there is no simple way to estimate the second-order exponent $\zeta_2$ from
the spectral exponent $\delta$ for such mildly active suspensions.
Nevertheless, it is simple, by using the procedure outlined above, to estimate
the $\zeta_p$ for such values of $\alpha$. In the lower inset of
Fig.~\ref{fig:zetap} (a), we present plots of $\zeta_p/\zeta_2$ vs. $p$, for $\alpha \gtrsim \alpha_c$,
which show that $\zeta_p/\zeta_2$ depends linearly on $p$ as noted for $0\leq p \leq 4$ in Ref.~\cite{wensink2012meso}. Therefore, in this mild activity regime there is no multiscaling.


\begin{figure}[h!]
	\includegraphics[width=1.0\columnwidth]{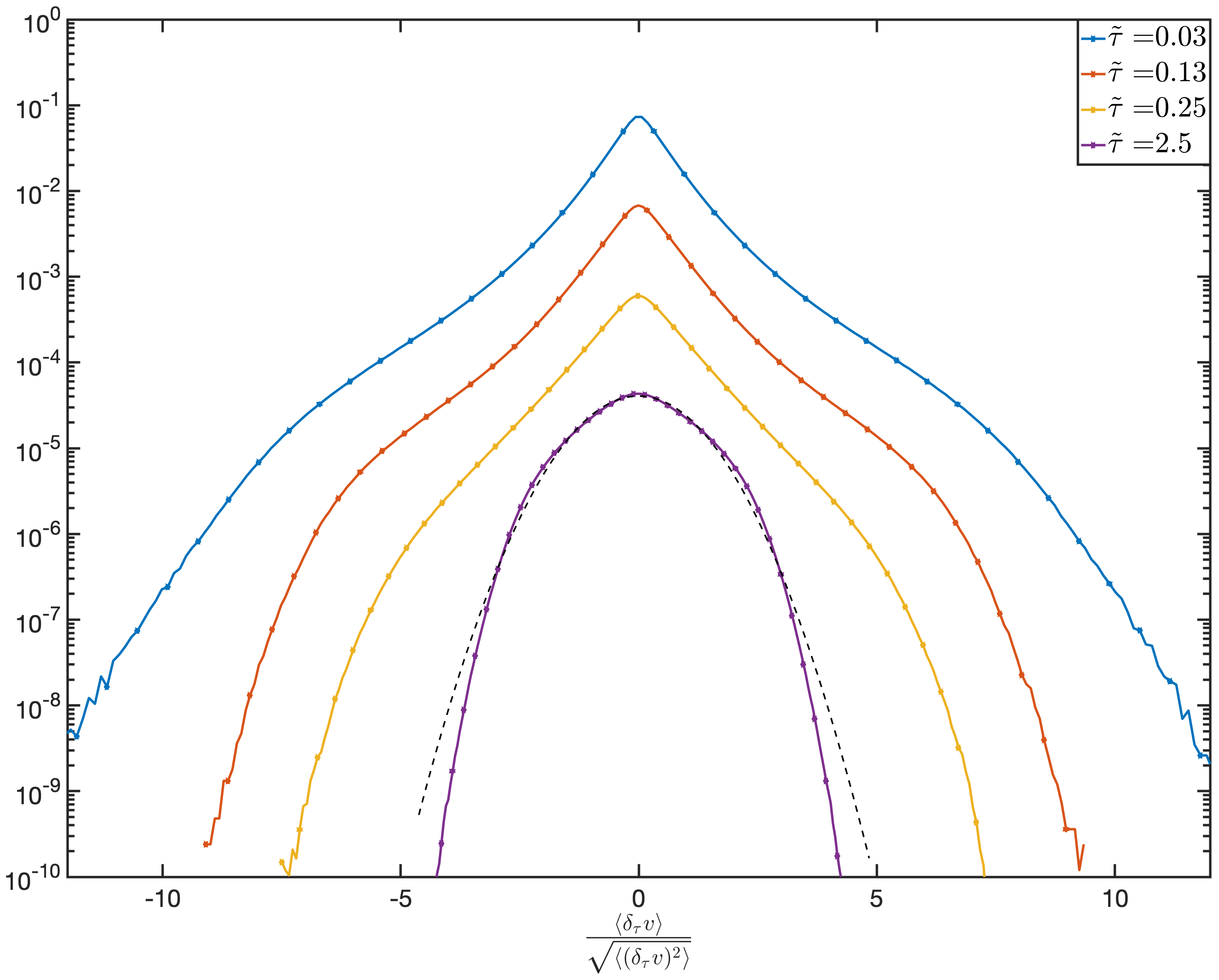}
	\caption{Probability density functions (PDFs) of the Lagrangian velocity increments $\Delta v_i$ (both Cartesian 
	coordinates included) for different time lags $\tilde{\tau}=\tau/(2\sqrt{2}\pi\Gamma_{2}/|\Gamma_{0}|^{2})$ (see legend) and for $\alpha = -6$. Clearly, these 
	distributions are Gaussian (indicated by the dashed line) at large time lags, but show significant non-Gaussian 
	tails as $\tau$ becomes smaller. (The curves are staggered vertically for clarity.)}
	\label{fig:pdf}
    \end{figure}

In inertial turbulence, intermittency manifests itself not only in 
to the Eulerian measurements, but also, and perhaps more crucially from the
point of view of transport, mixing, and dispersion, in Lagrangian studies. Thus, a full appreciation of this activity-induced, emergent
intermittency in low-Reynolds number flows requires us to investigate the
problem in the Lagrangian framework, where the governing equations for Lagrangian trajectories, $\bf x$, are given by
\begin{equation}{\label{eq:lag_evol}} 
\frac{d{\bf x}}{dt} = {\bf v};~{\bf v}({\bf x},t) = {\bf
u}({\bf x}).
\end{equation}
We measure, in particular, the Lagrangian
exponents $\xi_p$ via the Lagrangian structure functions $S_p^{\rm L} = \sum_i
\langle |\Delta v_i|^p \rangle \sim \tau^{\xi_p}$, where $\Delta v_i \equiv
v_i(t+\tau)-v_i(t)$, the time lag is $\tau$, and the index $i$ denotes the Cartesian
component. We focus first on highly active suspension, so we
present results for $\alpha = -6$. [We have checked that our results are robust
for all values of $\alpha \lesssim \alpha_c$.]

The PDFs of $\Delta v_i$ are instructive. For time lags much larger than the instability time scales, i.e., $\tau \gg 2\sqrt{2}\pi\Gamma_{2}/|\Gamma_{0}|^{2}$, these PDFs ought to be Gaussian (irrespective of the value of $\alpha$) as we see indeed 
in Fig.~\ref{fig:pdf}. However, for smaller values of $\tau \lesssim 2\sqrt{2}\pi\Gamma_{2}/|\Gamma_{0}|^{2}$, we show in Fig.~\ref{fig:pdf}
that these PDFs develop fat, non-Gaussian tails (for $\alpha \lesssim \alpha_c$), a clear indication of Lagrangian intermittency [cf. Refs.~\cite{mordant2001prl} and \cite{arneodo2008} for inertial turbulence]. We characterise these fat tails, below, by direct measurements of $\xi_p$. 

Before we turn to the actual measurements of the Lagrangian exponents, we recall that there is a simple dimensional
argument~\cite{biferale2004,arneodo2008} which leads to a relation between the Lagrangian exponents $\xi_p$ and 
the Eulerian exponents $\zeta_p$: If in time $\tau$, the Lagrangian particle moves from ${\bf x}_1$ to ${\bf
x}_2$, with ${\bf r} = |{\bf x}_2 - {\bf x}_1|$ and $r=|{\bf r}|$, and we assume that the corresponding Lagrangian-velocity
increment $\Delta v \sim |\Delta {\bf u}\cdot \hat {\bf r}| \sim S_1(r)$, then 
it follows that $\Delta v \sim r^{\zeta_1}$. The estimate $r \sim (\Delta v)\tau$ for
the distance traveled by the Lagrangian particle ,
yields $\Delta v \sim \tau^{\frac{\zeta_1}{1-\zeta_1}}$; this implies that $\xi_1 = \frac{\zeta_1}{1-\zeta_1}$.
If we assume that simple scaling holds (i.e.,  there is no intermittency), then $\zeta_p = p/4$; and $\xi_1 = 1/3$ and $\xi_p = p/3$. (For inertial turbulence, a similar argument gives $\xi_p = p/2$~\cite{biferale2004,arneodo2008}.)

To go beyond simple scaling, we must obtain the Lagrangian exponents $\xi_p$ 
from our DNS.
Once a statistically steady state is reached, we seed the flow, with $10^5$ tracer particles, initially
distributed randomly, and obtain their trajectories by integrating Eq.~\eqref{eq:lag_evol}. For time marching
we use a second-order Runge-Kutta scheme and use bilinear interpolation for $\bf{u}$ at off-grid points.

\begin{figure}
	\includegraphics[width=1.0\columnwidth]{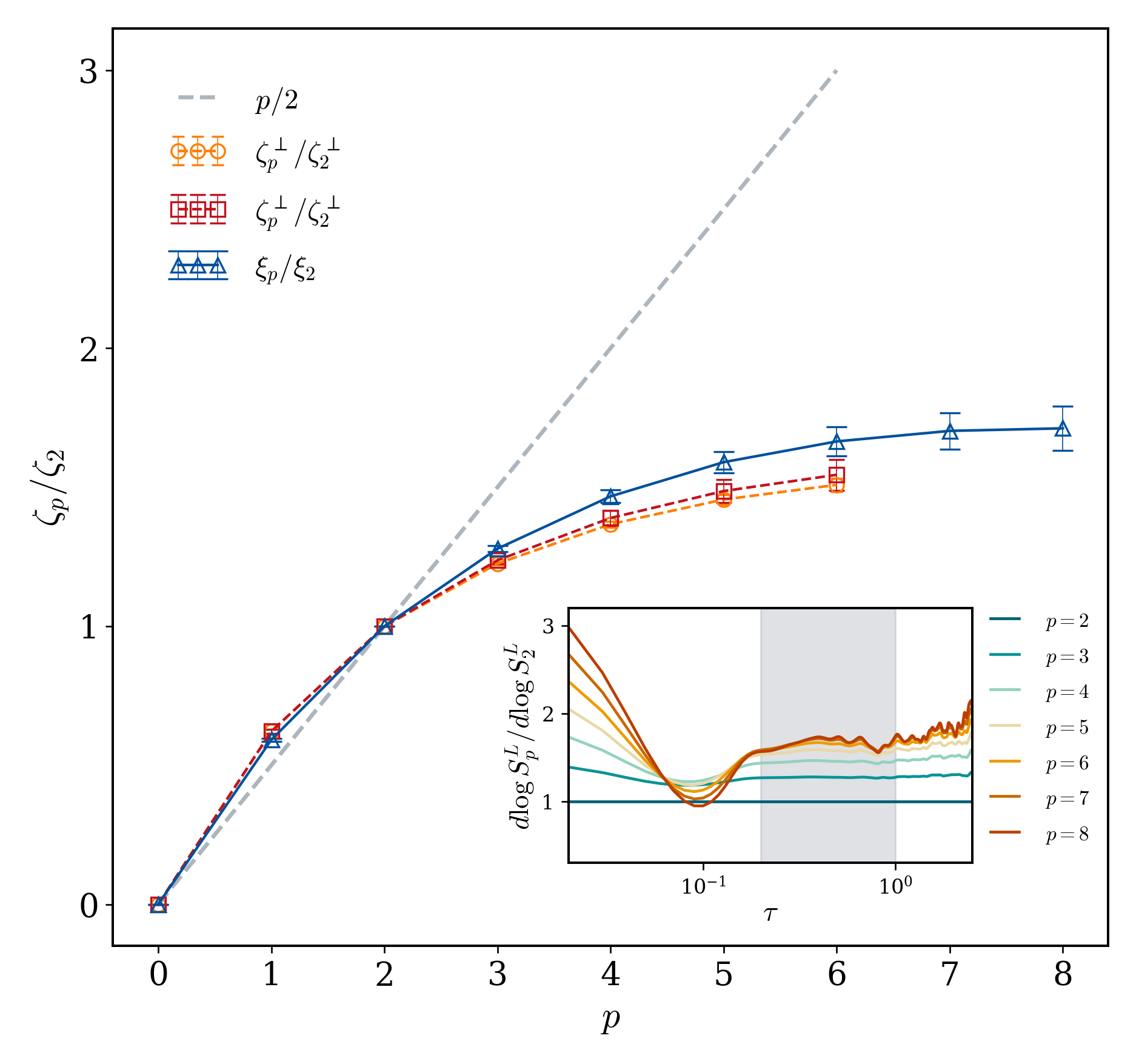}
	\caption{Plots (with error bars) of the Lagrangian exponent ratios $\xi_{p}/\xi_{2}$ and the Eulerian transverse structure function 
	exponent ratios $\zeta^{\perp}_p/\zeta^{\perp}_2$ (from two different simulations with different domain sizes; see legend) for $\alpha = -6$. 
	The solid black line gives the linear scaling result $p/2$. In the inset we show the local slopes $d \log S^{L}_{p}/d \log S^{L}_{2}$ versus the temporal increment $\tau$ for various orders. \textcolor{black}{The shaded gray region shows the plateau 
	whose mean value yields the scaling exponents; error-bars follow from the standard deviations.}}
	\label{fig:saturation}
    \end{figure}

We first focus on the second-order Lagrangian structure function to confirm $\xi_2 \simeq 2/3$.  
Then we use a combination of ESS~\cite{BenziESS,RayESS} and local-slope analysis to calculate
$\xi_p/\xi_2$.  In the inset of Fig.~\ref{fig:saturation}, we show
(for $\alpha = -6$) plots  of the local slopes $\frac{d \log S^{\rm L}_p}{d
\log S^{\rm L}_2}$ vs. $\tau$. In the limit $\tau \to 0$, the local slopes asymptote 
$\sim p/2$ beacuse small-scale smoothness yields the Taylor-expansion result $S^{\rm L}_p \sim
\tau^p$. For slightly larger values of $\tau \simeq \mathcal{O}(10^{-1})$, there is  a dip 
in the local slopes, which is familiar from similar measurements in inertial turbulence~\cite{arneodo2008,Buzzicotti_2016}),
where this dip is attributed to the trapping of tracers in vortices.
For even larger values of $\tau$, beyond this dip, a new scaling range emerges
for $S^{\rm L}_p(\tau)$, which corresponds to the scaling range for the Eulerian $S_p(r)$ [Fig.~\ref{fig:zetap} (c)], 
if we relate time and length scales via $\tau \sim r/\Delta v$. In the inset
Fig.~\ref{fig:saturation}, this range seen appears as a plateau after the
dip; the mean value of this plateau and the standard deviation from this mean yield the exponent ratios 
$\xi_p/\xi_2$ and their error bars, respectively. If there were no
intermittency, we would have found $\xi_p = p/3$ and $\xi_p/\xi_2 = p/2$. 
Our plot the exponent ratios as a function of $p$, in Fig.~\ref{fig:saturation},
shows significant deviations from the simple-scaling the $p/2$ line. This clear departure 
from linear scaling underlines the presence of Lagrangian intermittency in low-Re TTSH turbulence. 
Furthermore, we find another intriguing similarity with recent results for
inertial turbulence~\cite{iyer2020prf,BuariaPRL2023}: The exponents
$\xi_p$ saturate to a constant for $p \gtrsim 5$.
In inertial turbulence, this saturation of Lagrangian exponents has been attributed to the
saturation of the exponents $\zeta^{\perp}_p$ of the transverse, equal-time Eulerian
structure functions: $S^{\perp}_p (r) \equiv \langle |\Delta{\bf u}-(\Delta {\bf u}\cdot
\hat {\bf r})\hat{\bf r}|^p\rangle \sim r^{\zeta^{\perp}_p}$. We uncover a similar
correspondence with $\zeta^{\perp}_{p}\approx\xi_{p}$ for $p\gtrsim5$ (see
Fig.~\ref{fig:saturation}); these exponent ratios appear to saturate to $\simeq 1.5$. 
In inertial turbulence, it has been argued such saturation is likely because of 
slender vortex filaments~\cite{Frisch-CUP,BuariaPRL2023}. We conjecture that the exponent saturation we see, for $\alpha \lesssim \alpha_c$, may be related to localised streak-like structures, which have been seen in earlier studies~\cite{Sid21,Sid22,puggioni2022giant}.



We have also obtained $S^{\rm L}_p$ in the mild-activity range $\alpha_c < \alpha < 0$. We have shown above, the Eulerian structure functions $S_p$ display simple scaling in this range. By contrast, the behavior of $S^{\rm L}_p$ is more complex inasmuch as they show intermittent deviations from Gaussian statistics, but only at small values of $\tau$. However,
the Lagrangian structure functions $S^{\rm L}_p$ do not display a well-defined inertial range of scales, of the type we show in Fig.~\ref{fig:saturation} for the high-activity case $\alpha = -6$. 


\textcolor{black}{Our Eulerian and Lagrangian analyses of the emergent
multiscaling are interesting from the point of view of both turbulence and the
hydrodynamics of dense, active suspension.  They show that active turbulence in
these suspensions shares key fingerprints of inertial turbulence, namely, the
multiscaling of both Eulerian and Lagrangian structure function, in the
high-activity regime $\alpha \lesssim \alpha_c$. This emergent multiscaling, as
a function of $\alpha$, also resolves the contradiction in the findings from
experiments reported earlier.  In the experiments of Wensink \textit{et
al.}~\cite{wensink2012meso}, the variations in the velocity field are small and
the typical velocities $\simeq 25 \mu m/{\rm sec}$ lead to $\alpha \simeq -1 >
\alpha_c$. Hence, as reported by us in the lower inset of Fig.~\ref{fig:zetap}
(a), a simple scaling of $\zeta_p$ emerges. The typical large scale velocities in the experiments of Liu and I~\cite{liu2012multifractal} are similar to those in Ref.~\cite{wensink2012meso}, but show much strong variations in speeds, up to $\simeq 55 \mu
m/{\rm sec}$, leading to an effective $\alpha \simeq -4 \simeq \alpha_c$ \footnote{We refer the reader to the
Supplementary Information in Ref.~\cite{wensink2012meso} for details on the
conversion of experimentally measured velocities to the activity parameter
$\alpha$.}.
Indeed, their multiscaling exponents are consistent with Fig.~\ref{fig:zetap}
(a) and, in particular, with our measured $\zeta_2$. Significantly, these studies 
also complement, from a hydrodynamical approach, experimental results~\cite{secchi2016intermittent} suggesting a motility-induced transition 
to intermittent flows in \textit{B. Subtilis} suspensions.}

\textcolor{black}{How is multiscaling directly relevant to the individual active agents which make up our flow? 
It is tempting to conjecture that the 
the anomalous diffusion reported earlier for  $\alpha \lesssim \alpha_c$~\cite{Sid21,ariel2015swarming,gautam2024PRE} 
may well have its underlying roots in the Lagrangian multiscaling we report now. We do know 
that multiscaling in inertial turbulence has direct bearings on the motion of individual 
particles and in particular on how aggregates form~\cite{Becetal2016}. Emergent intermittent flow here could lead to similar advantages for individual agents in 
dense active suspensions.}

\textcolor{black}{Beyond possible biological implications of our results, there
is a second reason why this work is important. Our work advances
significantly the understanding of the dynamics of dense bacterial suspensions by isolating the truly \textit{turbulent} effects from those stemming from
simpler chaotic motion. More intriguingly, and at a broader conceptual
framework, this study yet again underlines that intermittency can be an
emergent phenomena in flows where the nonlinearity does not, trivially,
dominate the viscous damping.  Indeed, there is increasing evidence of
intermittency emerging in systems which are not turbulent in the classical
sense. Examples include flows with modest Reynolds number of $\sim
\mathcal{O}(10^{2})$ showing intermittent behaviour characteristic of high
Reynolds turbulence~\cite{schumacher2014pnas}, self-propelling active droplets
with intermittent fluctuations~\cite{nadia2023prr}, active matter systems  of
self-propelled particles, which undergo a glass transition, with an
intermittent phase before dynamical arrest~\cite{Mandal2020}, and perhaps most
pertinently, in elastic turbulence~\cite{Rahul1,Rahul2}.  Fundamentally, we present, for the first time, evidence of a critical threshold in
the control parameter $\alpha$ which allows a transition from a simple-scaling, non-intermittent flow to a multiscaling, intermittent one. 
Hence our studies throw up interesting questions
and offer a fresh perspective on understanding what causes flows to turn
intermittent and help in addressing fundamental questions in high-Reynolds-number inertial turbulence.}

\begin{acknowledgements}

	R. P. and K. V. K. acknowledge support from  the Anusandhan National Research Foundation (ANRF), the Science and Engineering Research Board (SERB), and the National Supercomputing Mission (NSM), India, for support, and the Supercomputer Education and Research Centre (IISc), for computational resources. KK and SSR are indebted to S. Mukherjee for his tutelage on understanding active turbulence as well as developing the DNS code at ICTS.  The CPU/MPI	simulations were performed on the ICTS clusters \emph{Tetris} and
	\emph{Contra}. The GPU simulations were performed on the CUDA A100 at Departement of Physics, IISc. KVK and SSR thank Rahul Singh for careful reading of the manuscript. SSR acknowledges SERB-DST (India) projects
	STR/2021/000023 and CRG/2021/002766 for financial support. KK and SSR
	acknowledges the support of the DAE, Govt. of India, under project no.
	12-R\&D-TFR-5.10-1100 and project no. RTI4001. AG acknowledges SERB-DST (India) projects MTR/2022/000232, CRG/2023/007056-G for financial support. This research was supported in part by the International Centre for Theoretical Sciences (ICTS) for the discussion meeting- Turbulence and Vortex dynamics in 2D quantum fluids, code: ICTS/QUFLU2024/02.

\end{acknowledgements}
%
\end{document}